\documentclass[oribibl]{llncs}

\usepackage{amsmath}
\usepackage{caption}
\usepackage{subfigure}
\usepackage{graphicx}
\usepackage{mathtools}
\usepackage{array}
\usepackage{multirow}
\usepackage{ amssymb }

\newcommand{\BigO}[1]{\ensuremath{\operatorname{O}\bigl(#1\bigr)}}

\newtheorem{mydef}{Theorem}

\begin{document}

\title{Computing Real Numbers using DNA Self-Assembly}

\author{Shalin Shah, Parth Dave and Manish K Gupta\\  
Email :- \{shah\_shalin, dave\_parth, mankg\}@daiict.ac.in}

\institute{Laboratory of Natural Information Processing, \\
Dhirubhai Ambani Institute of Information and Communication Technology, \\
Gandhinagar 382007, Gujarat, India
} 

\maketitle
\bibliographystyle{IEEEtran}

\begin{abstract}
DNA Self-Assembly has emerged as an interdisciplinary field with many intriguing applications such DNA bio-sensor, DNA circuits, DNA storage, drug delivery etc. Tile assembly model of DNA has been studied for various computational primitives such as addition, subtraction, multiplication, and division. Xuncai et. al. gave computational DNA tiles to perform division of a number but the output had integer quotient ~\cite{XuncaiSubDiv}. In this work, we simply modify their method of division to improve its compatibility with further computation and this modification has found its application in computing rational numbers, both recurring and terminating, with computational tile complexity of $\BigO{1}$ and $\BigO{h}$ respectively. Additionally, we also propose a method to compute square-root of a number with computational tile complexity of $\BigO{n}$ for an n bit number. Finally, after combining tiles of division and square-root, we propose a simple way to compute the ubiquitously used irrational number, $\pi$, using its Gregory–Leibniz infinite series. 
\end{abstract}

\section{INTRODUCTION}
In the last two decades, DNA has been used for archival data storage~\cite{DNACloud}, molecular logic circuits~\cite{QianCircuits},  building complex 2D structures from single stranded DNA~\cite{PengOrigami} as well as synthetic 3D polyhedra~\cite{PengPolyhedra}. This high utility of DNA is because of its known nano-scale structure, programmable nature, and  flexibility~\cite{ReviewDNAStructure}. Using double crossover (DX) and triple crossover (TX) molecules, Winfree showed that DNA self-assembly is Turing Universal~\cite{WinfreeThesis, WangWork, SeemanWork, SeemanTiling}. To approximate self-assembly of DNA, he also proposed two models - abstract Tile Assembly Model (aTAM) and kinetic Tile Assembly Model (kTAM). Many well-known problems have been solved using tile assembly model~\cite{AdelmanHamiltonPath, IEEEKnapsack, BrunSubsetSum, YuriAddMul, BrunFactor}. In particular, tile-assembly models to perform arithmetic computations such as addition, multiplication and factor were shown by Yuri Burn~\cite{BrunSubsetSum, YuriAddMul, BrunFactor}. Addition and multiplications operations had computational tile complexity of \BigO{1} and running complexity of \BigO{n}~\cite{YuriAddMul} where as the factoring numbers nondeterministically had $\BigO{1}$ distinct components~\cite{BrunFactor}. Similarly, Xuncai et. al. also came up with computational tiles for subtraction and division of integers, again, using \BigO{1} computational tiles. The running time complexity is \BigO{n} for subtraction and \BigO{n^{2}} for division~\cite{XuncaiSubDiv}. Additionally, open source softwares such as XTile (it can convert any computational tile formula to a .tile file supported XGrow~\cite{WinfreeXGrow}), ISU TAS~\cite{ISUTAS} and XtileMod (it can generate required computational tiles for performing arithmetic operations as well as the corresponding .tiles file for DNA tiles) have been developed~\cite{XTile, XTileMod}. 

In the present paper, we propose a method to compute rational number, irrational number($\pi$), and square-root of a number. Additionally, we also propose generalized method to compute division by extending Xuncai's method. In \cite{XuncaiSubDiv}, the output of the division was in the form of quotient and remainder which, however, we have changed to decimal form. This paper could be useful for creating DNA based calculator.

The paper is organized as follows. In Section $2$, we briefly introduce aTAM. Section $3$ provides basic computational tiles of Xuncai et. al. division. This tile systems are also used in computing square-root, rational number and $\pi$ with some minor modifications, in subsequent sections. In Section $4$, we provide computational tiles to perform division in rational form. In Section $5$, we have given an interesting additional tile system, the insertion tile system, required to compute square-root. Also, we show how to compute square-root. In Section $6$, we have given method to compute rational number and $\pi$ as application of the square-root tile system. In Section $7$, we briefly tell about our command line tool which can be used to generate .tile file for computing square-root of a number. Section $8$ concludes paper with general remark.

 \section{BACKGROUND}
In \cite{WinfreeThesis}, Winfree gave abstract Tile Assembly Model (aTAM) and kinetic Tile Assembly Model (kTAM) to understand the relation between self-assembly and computation. The basic building block of aTAM is a unit square tile. The edges of these tiles are nothing but sticky ends of DNA which can combine with other sticky ends of same edge label or edge color. Every tile has $4$ edges - North, South, East and West - with a corresponding edge color and glue strength. Also, it is to be noted that, a tile cannot be rotated. In this paper, most of the symbols used have identical meaning to those used in \cite{XuncaiSubDiv}. Mathematically, a tile system is represented as $S_R = (T, S, g, \tau)$, as shown in Fig.~\ref{imageSelfAssembly}, where \textbf{T} is the tile system, \textbf{S} is the seed configuration of this tile assembly, \textbf{g} is the glue strength of an edge, and \textbf{$\tau$} (always $\ge 0$) is the threshold temperature. The tile assembly starts growing from seed tile and input tiles at the bottom of the tile assembly system. A new tile will attach to the current tile if :

\begin{figure}[ht!]
\centering
\includegraphics[scale=0.4]{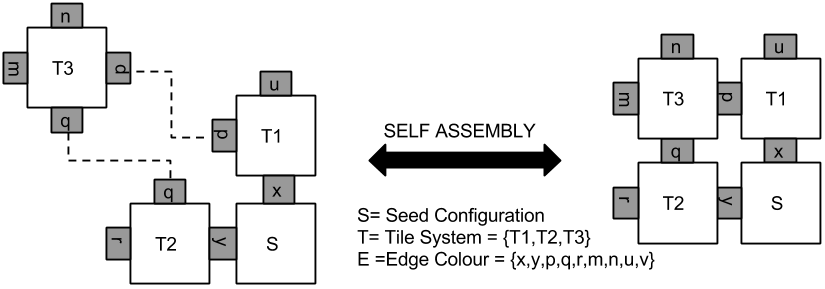}
\caption{The image shows basic concept for self-assembly of tiles.}
\label{imageSelfAssembly}
\end{figure}

\begin{itemize}
\item Edges have same glue color or glue label.
\item The combined glue strength of both the tiles is greater than or equal to the threshold temperature ($\tau$). 
\end{itemize}

For a given a seed configuration, we can represent above mentioned conditions in mathematical form. A tile t can attach at point (x, y) if following condition is satisfied~\cite{XuncaiSubDiv}:  

\begin{equation}
\sum_{d \epsilon D} g(bd_D(t), bd_{D^{-1}}(S_R(x, y))) \ge \tau
\end{equation}

Here, D indicates set of four directions i.e \{N, E, S, W\} and $D^{-1}$ indicates direction opposite to D. $bd_D(t)$ indicates binding domain of tile t in direction D. g(a, b) indicates combined glue strength of glue colors a and b. Given this set of condition, a tile assembly will start growing from initial tile set configuration. Boundary tiles with one edge of glue strength zero are used to avoid any further growth of tiles. We represent a number \textit{a} = $\sum\limits_{i = 0}^{n - 1} {a^i 2_i}$ by $a_0a_1a_2a_3\dots a_{n-1}$ where $a_0$ is the most significant bit and $a_{n-1}$ is the least significant bit. For more details, the reader should refer to \cite{XuncaiSubDiv, WinfreeThesis}.

\section{TILE SYSTEMS}
Square-root is function which is provided, almost, by every existing calculator. However, in order to compute square-root using DNA tiles, we propose to use four different kinds of tiles systems, which are similar to those provided by \cite{XuncaiSubDiv}, however, with some modifications. The similar tile systems will also be used to compute rational numbers and $\pi$.

\begin{enumerate}
  \item Comparator Tile System
  \item Shift Tile System
  \item Shift and Subtract Tile System
\end{enumerate}

\subsection{COMPARE TILE SYSTEM}
\begin{figure}[ht!]
\centering
\includegraphics[width=0.8\textwidth]{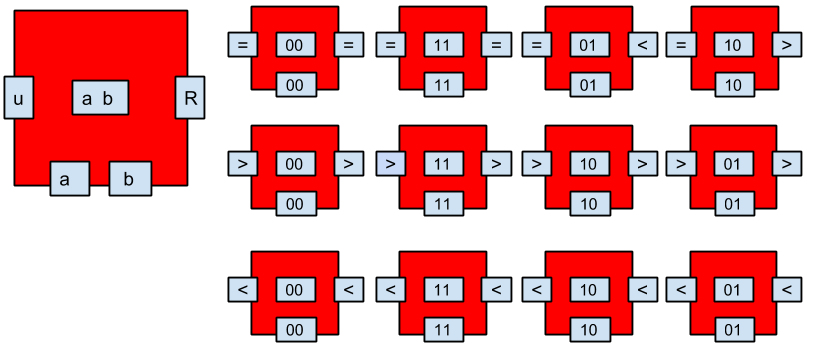}
\caption{Comparator System: Left Tile is the general tile representation for all the $12$ comparator tiles. Here a, b, c $\epsilon \{0, 1\}$.}
\label{imageCompare}
\end{figure}

\begin{figure}[ht!]
\centering
\includegraphics[width=0.8\textwidth]{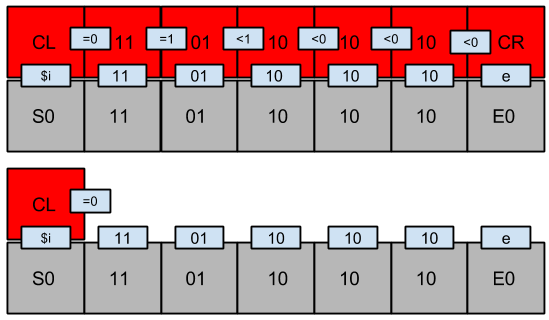}
\caption{The image shows how compare tile system grows from seed configuration for a = $10111_{2}$ and b = $11000_{2}$.}
\label{imageCompareExample}
\end{figure}

Compare tile system~\cite{XuncaiSubDiv} helps us to identify relationship between two numbers a and b that is whether a is \{$>$, $<$ or $=$\} than b. As shown in the Fig.~\ref{imageCompare}, a basic comparison tile consists of $12$ different tiles where \textbf{\textit{R}} = \{$<$, $>$ or $=$\} depending on the relationship between $a_0a_1a_2 \ldots a_{k-1}$ and $b_0b_1 \ldots b_{k-1}$. For example, if a = $10111_{2}$ and b = $11000_{2}$ then, as shown in Fig.~\ref{imageCompareExample}, we take input tiles, S0, and CL in seed configuration which grows to unique configuration as shown in Fig.~ \ref{imageCompareExample}. Here, since $a_0=b_0$ so \textbf{\textit{R}} = \{$=$\} for first bit. For second bit, $a_{1}<b_{1}$ so \textbf{\textit{R}} = \{$<$\}. Once we get a \{$>$ or $<$\}, we continue this sign till the end of system since irrespective of coming bits since the first number is already proved \{$>$ or $<$\} than other. This proved that a $<$ b so all the succeeding tiles will have same \textbf{\textit{R}}.  The proof that compare tile system is a step configuration, total k +1 steps so \BigO{k}, and producer of unique tile system for the given seed configuration has already been given in \cite{XuncaiSubDiv}. 


\subsection{SHIFT TILE SYSTEM}
\begin{figure}[ht!]
\centering
\includegraphics[width=1\textwidth]{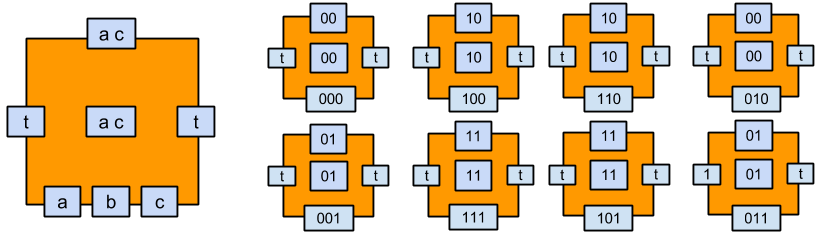}
\caption{Shift System: Left Tile is the general tile representation for all the $8$ shift tiles.}
\label{imageShift}
\end{figure}

\begin{figure}[ht!]
\centering
\includegraphics[width=0.8\textwidth]{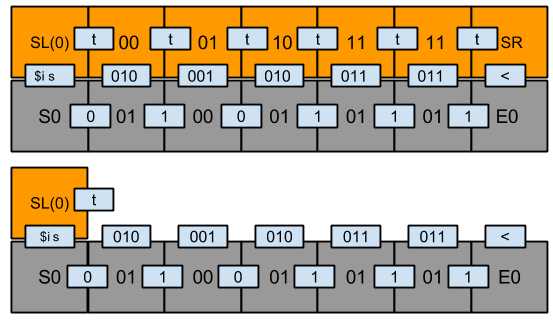}
\caption{The image shows how shift tile system grows from seed configuration for a=$10111_2$. The number a, which is to be shifted, is right bit of input tiles.}
\label{imageShiftExample}
\end{figure}

Shift tile system~\cite{XuncaiSubDiv} helps us to right shift the bits of a k bit number by one place while also padding a 0 in most significant bit. As shown in Fig.~\ref{imageShift},  the system has input bit a, b, and c coming from the south end of tile. Here, a and b represent the input bits and c represents the right bit of previous tile (left tile). For example. as shown in Fig.~\ref{imageShiftExample}, for input a = $10111_2$, seed configuration includes S0, SL, and input tiles. This seed configuration gives unique final output a = $01011_2$, as shown in Fig.~\ref{imageShiftExample}, where a is shifted right by one bit. The proof for the shift tile system is given in \cite{XuncaiSubDiv} as the Duplicator Tile System. This seed configuration also produces unique final output and is a step configuration with time complexity \BigO{k} for a k bit input number. 

\subsection{SUBTRACT \& SHIFT TILE SYSTEM}
\begin{figure}[ht!]
\centering
\includegraphics[width=0.87\textwidth]{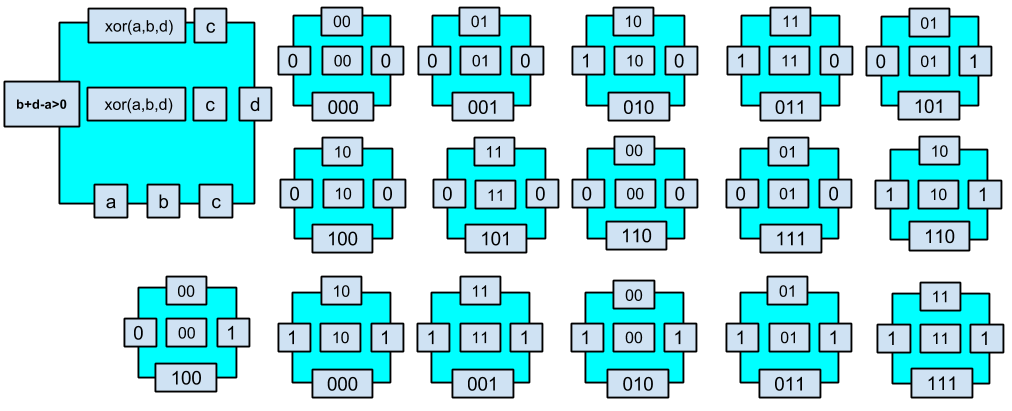}
\caption{Shift and Subtract System: Left Tile is the general tile representation for all the $16$ shift tiles.}
\label{imageSubShift}
\end{figure}

\begin{figure}[ht!]
\centering
\includegraphics[width=0.75\textwidth]{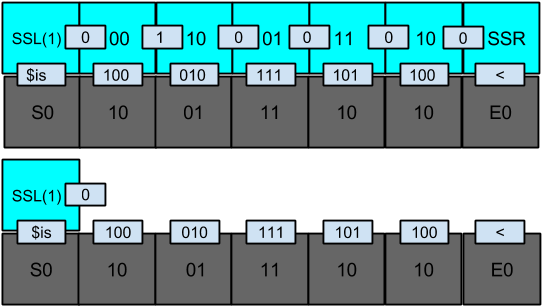}
\caption{The growth of subtract and shift tile system from seed configuration for a=$10111$ and b=$01100$.}
\label{imageSubShiftExample}
\end{figure}

Subtract \& shift tile system~\cite{XuncaiSubDiv} is used to subtract right input bits (divisor) from left input bits (divident) and then shift the right bits by one. Fig.~\ref{imageSubShift} shows computational tiles for subtracting and shifting a number. Here, a, b, c, and d are inputs where b is current right bit and c is the right bit of previous tile. This c bit, therefore, helps to right shift the number and xor operation helps in subtraction. For example, as shown in Fig.~\ref{imageSubShiftExample}, let a=$10111$ and b=$01100$. Hence, growth of this tile system gives a - b = 01011 as left bits and right sift version of b as right bits. Again, the proof and details of this method are provided in \cite{XuncaiSubDiv}, so we will not go in to details here. But this tile system is also step configuration with time complexity of \BigO{k} taking total of $k + 1$ steps for a k bit input number. 

\section{EXTENDING DIVISION}
\begin{figure}[ht!]
\centering
\includegraphics[width=0.65\textwidth]{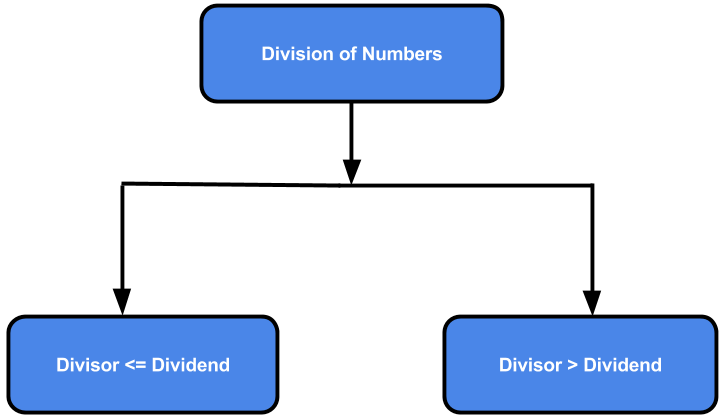}
\caption{Flow chart showing both the cases for computing division of a number.}
\label{imageDivisionCases}
\end{figure}

Earlier, Xuncai et. al. had proposed a method to compute division using DNA tiles~\cite{XuncaiSubDiv}. However, their idea of division considers dividing dividend by divisor to get quotient and remainder. Instead, here we propose a method to get output in different form. We give quotient in decimal form so that one can use this decimal quotient, subsequently, in doing further calculations using tile assembly. Computing division is divided in two cases as shown in the Fig.~\ref{imageDivisionCases}.
 
\begin{figure}
\begin{minipage}{.5\textwidth}
  \centering
  \includegraphics[width=1\linewidth]{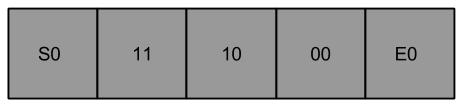}
  \captionof{figure}{Divident = 6  and divisor = 1.}
  \label{imageDivisionCase1}
\end{minipage}
\begin{minipage}{.5\textwidth}
  \centering
  \includegraphics[width=1\linewidth]{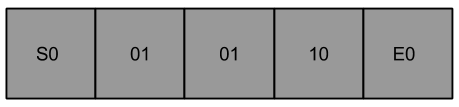}
  \captionof{figure}{Divident = 1 and divisor = 6.}
  \label{imageDivisionCase2}
\end{minipage}
\end{figure}

We have to see whether divisor is greater than dividend or vice versa. If the divisor is greater than dividend then we write dividend bits from its right end and prepend zero's, as shown in Fig.~\ref{imageDivisionCase2}. In the other case, we write divisor bits from its left end and append zero's for the remaining bits, as shown in Fig.~\ref{imageDivisionCase1}. In the paper, we give an example of division where quotient turns out to be infinite. 

\begin{figure}[ht!]
\includegraphics[scale=0.35]{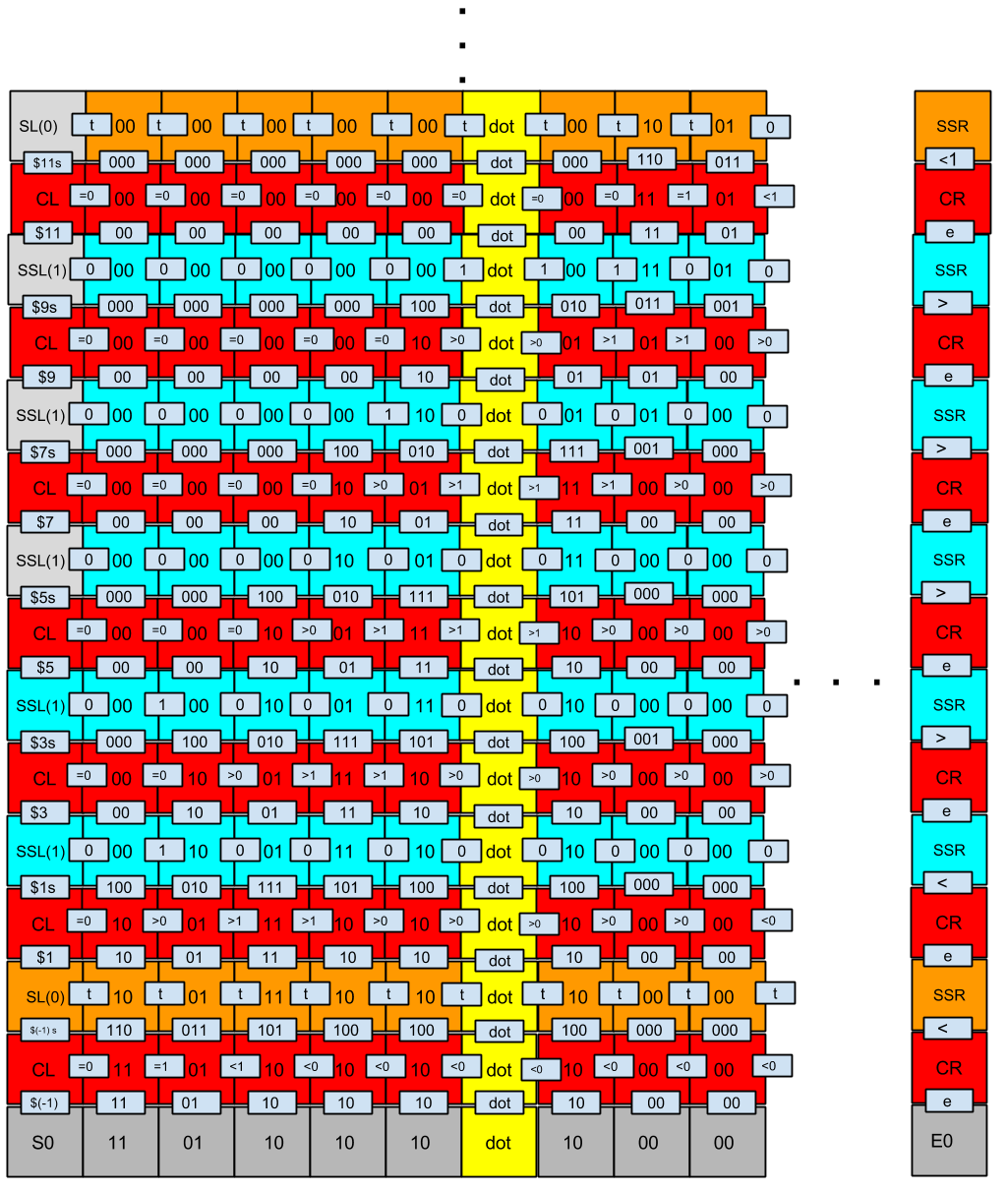}
\caption{Division of a = 23.5 by b = 6 is calculated by this self-assembly. Here a is encoded in binary as $10111.1_{2}$ and b is encoded as $110_2$. We get $11.1110_{2}$ as the output at the left side of the tile assembly model. Yellow tiles simply indicates decimal point. Grey tiles indicate values after decimal point.}
\label{imageDivision}
\end{figure}

Fig.~\ref{imageDivision} shows the tile assembly growth for division of 23.5 by 6. The answer to this division is $3.91666666\ldots$ - a recurring rational number. Suppose we want first eight seven bits of the number. In that case, we have to give eight bit input to the tile assembly so that we get 5 bits before decimal place and 3 bits after decimal place. Note that this division is an example of case 1 as depicted in Fig.~\ref{imageDivisionCase1} but the number of bits are, inadvertently, same for both divisor and dividend.

Note that the \textit{dot} tiles used in the division process are optional. They are just for better understanding of the reader. Also, the square dots in the Fig.~\ref{imageDivision} indicate that the assembly can be extended further if we want more answer bits. 

The SSL and SL bits represent our answer - quotient - when seen in bottom-up fashion. In this case, we connect $0111110\ldots _2$ from the left boundary tiles and we place a decimal point after three bits which gives us  $011.1110_2$ = $3.916$. After we get eight bits of quotient, we can terminate growth of this tile assembly with the help of boundary tiles. To sum up, we have not used any additional tiles to compute quotient in decimal form that is these are the exact same tile used in \cite{XuncaiSubDiv}. Thus, the computation uses $\BigO{1}$ computational tiles and this computation takes $\BigO{k^2}$ steps for a k bit number. However, if we wish to terminate the computation then we will require $\BigO{h}$ boundary tiles as shown in Fig.~\ref{imageDivision} where h is the height of assembly.

\section{SQUARE-ROOT}
In this section, by combining the tile systems mentioned in Section 4, we illustrate how to find square-root a number. We will describe a new tile system required to compute square-root of a number.

\subsection{INSERT BIT TILE SYSTEM}
\begin{figure}[ht!]
\centering
\includegraphics[width=0.5\textwidth]{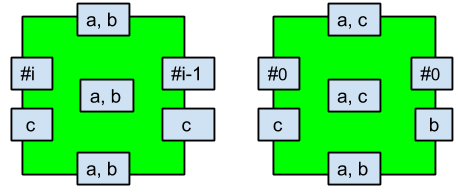}
\caption{Insert Tile System: General Tiles for inserting a bit at $i^{th}$ position from left side. Here a, b, c $\epsilon \{0, 1\}$ and $1 \le i \le n-1$.}
\label{imageInsert}
\end{figure}

\begin{figure}[ht!]
\centering
\includegraphics[width=0.75\textwidth]{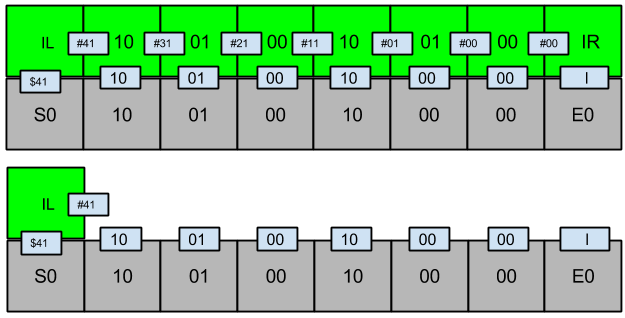}
\caption{The assembly inserts $1$ at the $5^{th}$ position from left in a = $100100_2$.}
\label{imageInsertExample}
\end{figure}

This tile system is used to insert a bit in a number. Fig.~\ref{imageInsert} shows the computational tiles required to do this operation. As shown in the Fig.~\ref{imageInsert}, \textit{a}, \textit{b}, \textit{c}, and \textit{\#i} are the inputs. The values on the east side are output values so this tile system grows from west to east. Here, we subtract the value of {\textit{i}} unless it becomes zero and as soon at it does, we stop. We do this since we want to insert bit at $i^{th}$ position in the number. So, as soon as \textit{i} becomes zero, we place the value of \textit{c} there and after this tile, we shift all the remaining bits by one to right. For example, as shown in Fig.~\ref{imageInsertExample}, to insert 1 at $5^{th}$ position in $100100_2$, we take \#41 as the east input of IL which is reduced unless 4 becomes 0 to insert 1 as right bit at that position and shift later bits by one place. 

Let us take an example number n = $100100_{2}$. Suppose, we want to insert number $1$ at fifth position. Therefore, as shown in Fig.~\ref{imageInsertExample}, $bd_N(S(-k, -1)) = (\#4, 1)$.  Now, in the second level, we keep on reducing the \#i value unless we get zero. As soon as we get a zero, we insert the \textit{c} bit at that position. After this point, we simply right shift the remaining values.

\begin{mydef}
Let $\sum$ be an alphabet defined as

\begin{tabular}{ l l r }
  $\sum =$  & $\{00,01, 10, 11, l\}$ & $\bigcup$ \\
  & $\{\#i c | 1 \le i \le n - 1$ and i = 2k + 1  ; c $\in \{0, 1\}\}$  \\
\end{tabular}

.\\and $T_1$ be the set of tiles defined over $\sum$. See Fig.~\ref{imageInsert}. let $g=1$, $\tau=2$, and S be a seed configuration as defined in Fig.~\ref{imageInsertExample}. Let q be the number to insert the bit and let m and n be the sizes, in bits, of p and q, respectively. If $m<n$, the number q needs to be padded to be m bits long with extra 0 in the $q’s$ high bit. Let $S_R=(T_1, S, g, \tau)$. Then, there exists some $(x_0, y_0) \epsilon Z_2$ such that: $S(x_0 + 1, y_0 - 1) = E_0, S(x_0 - k, y_0 - 1) = S_0, S(x_0 - k, y_0) = IL, S(x_0 + 1, y_0) = IR$; for all $0 \le i \le k-1, bd_N(S(x_0 - i, y_0 - 1)) = p_iq_i$. For all other positions (x, y), S(x, y) = empty. Then, S produces a unique final configuration F on S to insert a bit at $i^{th}$ position in q. The assembly configuration time is \BigO{m}.
\end{mydef}

\textbf{Proof.} Consider the tile system $S_R$. Let input tiles be $T_1$ = $\{E_0 = \{e, null, null, null\}, \\
D_{00} = \{00, null, null, null\}, I_L = \{null, \#ix, \$ ix, null\}, D_{01} = \{01, null, null, null\}, \\
D_{10} = \{10, null, null, null\}, D_{11} = \{11, null, null, null\}, S_0 = \{\$ix, null, null, null\}$\}. \\

Let the seed configuration $S: Z_2 \rightarrow T_1$ be:
\begin{equation}
 \begin{cases}
    E_0 = S_R(1, -1)\\
    S_0 = S_R(-k, -1)\\
    IL = S_R(-k, 0)\\
    \forall i \epsilon \{0, 1, .., k-1\}, S_R(-i, -1) = D_{pq}\\
    $For all other (x, y)$ \epsilon Z_2, S_R(x, y) = empty
  \end{cases}
\end{equation}
For the given seed configuration $S$ as shown in Fig.~\ref{imageInsertExample}, there is only one position where the incoming tile can attach. Also, given that $\tau=2$, by induction we can say that $\forall t \epsilon$ $T_1$,  $(bd_S(t), bd_W(t))$ is unique. Therefore, it follows that $S_R$ produces unique final configuration $F$ and it takes $k+1$ steps to assemble.

As shown in the Fig.\ref{imageInsert}, value of \#i indicates whether the bit needs to be inserted or simply copied to the other side. If (\#i, c) has $i > 0$ then the output on east will be (\#(i - 1), c). If, on the other hand, we have (\#0, c) as the input on west side, then we will replace (a, b) by (a, c) and the output on east side will be (\#0, b).

Let the final configuration be F. Also, S and F agree on the points (-i, -1), (-k, -1) and (-k, 0) for all $0 \le i \le k - 1$. Therefore, $bd_N(F(-i, -1)) = pq$. From the definition of $S_R$, $bd_W(F(-k, 0))=(\#i, c)$ and the $bd_W(F(-(k-1), 0))=(x, y)$ where y depends on the value of x. If $\exists m \le k - 1$ where \#i = \#0, and for all $n > m$, $i > 0$, then for all $m >n$ we have $bd_E(F(-n, 0)) = (\#(i -1), c)$ and for all $0 \le n \le m$, we have $bd_E(F(-n, 0)) = (\#0, b)$ and $v(-n, 0) = (x, c)$. After the relationship is determined and right boundary tile IR is attached at (1, 0), the insertion process will terminate. Additionally, we have $\tau = 2$ which means that only tile will attach at a time; the one which has two neighboring tiles. To sum up, this clearly indicates that we will have unique tile configuration which takes $k + 1$ steps to finish $\blacksquare$
  
\subsection{SQUARE-ROOT ASSEMBLY}
\begin{figure}[ht!]
\centering
\includegraphics[width=0.85\textwidth]{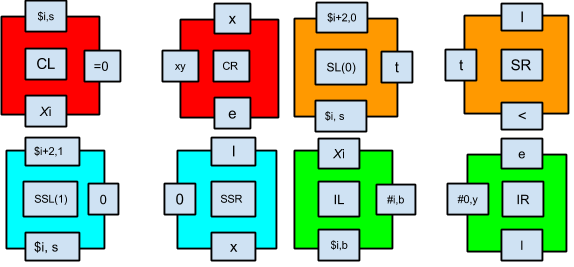}
\caption{Boundary tiles for compare, shift, insert, and subtract \& shift (clockwise). Here, the variables used in the glue colors of boundary tiles are :
$\{\$i x | -1 \le i \le n - 1 ;  x \in \{0, 1, s\}\}$, $\{Xi  | 1 \le i \le n - 1$; $i = 2k + 1\}$, \\
$\{\#i b | 1 \le i \le n - 1 ;  b \in \{0, 1\}\}$, $\{xy |   x \in \{<, >, =\}; y \in \{0, 1\}\}$, $\{x | x \in \{>, =\}\}$.}
\label{imageBoundary}
\end{figure}

\begin{figure*}[ht!]
\includegraphics[scale=0.4]{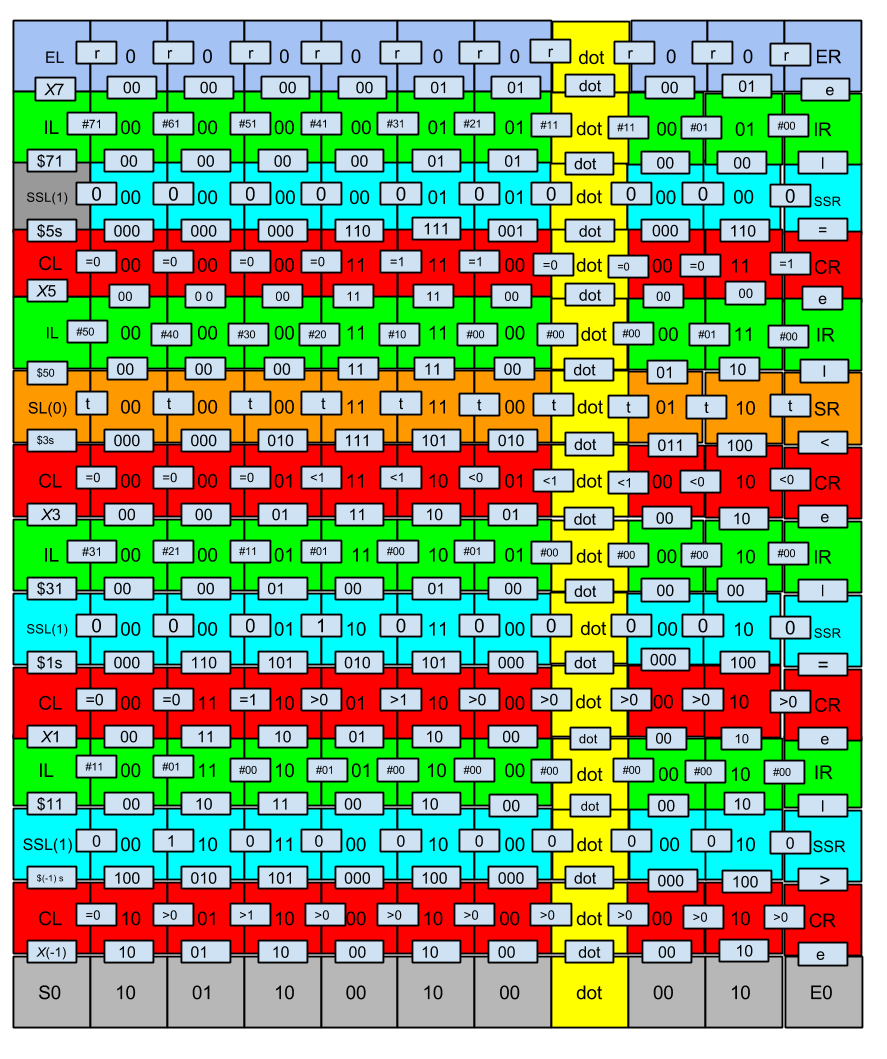}
\caption{Square Root of n = 42.25 is calculated by the self-assembly of tiles. Here n is encoded in binary as $101010.01_{2}$. We ge $110.1_{2}$ as the output at the left side of the tile assembly model. Color of the tile represent same tile-set as mentioned. Yellow tiles simply indicates decimal point. Grey tiles indicate values after decimal point.}
\label{imageSquareRoot}
\end{figure*}

Suppose we wish to find square-root of a number then we write its bits on left side in input tiles. If the length of n is odd then we add an extra zero bit in the leftmost tile and then start filling the bits. The right bits of the input tile system has 0 and 1 in the two leftmost tiles respectively and all other tiles have zero in the right part. Fig.~\ref{imageSquareRoot} show the input configuration for n = 42.25. In this system, there are four operations namely compare (red tiles), shift (orange tiles), subtract and shift (blue tiles), and insert bit (green tiles). Compare and Insert operations have assembly growth from west to east while the others have it from east to west. As shown in Fig.~\ref{imageSquareRoot}, we compare input numbers, p = $101010.01$ and q= 010000.00, from west to east and since p$>$q subtract and shift tile system attaches. After the growth of this tile system, we insert 1 at $2^{nd}$ position from left. Now, the process is repeated that is compare, shift or subtract and shift, and insert perpetually unless the northern glue of insertion tiles becomes equal to \$(n-1) and then the termination end tiles self assemble themselves. At any point if compare tile give $<$ as output then instead of subtract and shift boundary tile, shift boundary tile will attach to CR. The boundary tiles for all these tile systems are shown in Fig.~\ref{imageBoundary}. 

Note that by rational number we mean that the number should be recurring or terminating. However, in the recurring case, like calculator, we take first 8 bits after decimal. 
 
\begin{mydef}
Let $\sum$ be an alphabet defined as

\begin{tabular}{ l l r }
  $\sum =$  & $\{0, 1, 00, 01, 10, 11, e, l, t,  <, >, =, r, dot \}$ & $\bigcup$ \\
  & \{xy $|$ x $\epsilon\{<, >, =\};$ y $\epsilon \{0, 1\}\}$ & $\bigcup$ \\
  & $\{000, 001, 010,  011, 100, 101, 110, 111\}$ & $\bigcup$ \\
  & $\{\$i  | -1 \le i \le n - 1$; $i = 2k + 1\}$ & $\bigcup$ \\
  & $\{Xi  | 1 \le i \le n - 1$; $i = 2k + 1\}$ & $\bigcup$ \\
  & $\{\#i b | 1 \le i \le n - 1 ;  b \in \{0, 1\}\}$ & $\bigcup$ \\
  & $\{\$i x | -1 \le i \le n - 1 ;  x \in \{0, 1, s\}\}$ 
\end{tabular}
\\\.\\and $T_2$ be the tile set defined over $\sum$. SeeTable \ref{tileTable}, given that $\tau = 2$ and g=1, can compute square root of a n bit rational number defined as  $S_R = (T_2, S, g, \tau)$ where S is the seed configuration.. Then, there exists some $(x_0, y_0) \epsilon Z_2$ such that : $S_R(x_0 - k, y_0 -1) = S_0$, $S_R(x_0, y_0 - 1) = E_0$, $S_R(x_0 - k, y_0) = CL$, and for all i $\epsilon \{0, 1,...,k-1\}$, $bd_N(x_0 - i, y_0 - 1) = pq$. Then, the configuration S can produce unique final configuration F and will compute square-root of a number.
\end{mydef}

\textbf{Proof.} Consider the tile system $S_R$. Let input tiles be $T_2$ = $ E_0 = \{e, null, null, null\},\\
D_{00} = \{00, null, null, null\}, D_{01} = \{01, null, null, null\}, \\ 
 D_{10} = \{10, null, null, null\}, D_{11} = \{11, null, null, null\}, S_0 = \{\$ix, null, null, null\}\}$. \\

Let the seed configuration $S: Z_2 \rightarrow T_2$ be:
\begin{equation}
 \begin{cases}
    E_0 = S_R(1, -1)\\
    S_0 = S_R(-k, -1)\\
    CL = S_R(-k, 0)\\
    \forall i \epsilon \{0, 1, .., k-1\}, S_R(-i, -1) = D_{pq}\\
    $For all other (x, y)$ \epsilon Z_2, S_R(x, y) = empty
  \end{cases}
\end{equation}
For the given seed configuration $S$ as shown in Fig.~\ref{imageInsertExample}, there is only on position where the incoming tile can attach. Also, given that $\tau=2$, by induction we can say that $\forall t \epsilon$ $T_2$,  $(bd_S(t), bd_W(t))$ and $(bd_S(t), bd_E(t))$ is unique. Therefore, it follows that $S_R$ produces unique final configuration $F$ and it takes $k+1$ steps to assemble.

Let the final configuration be F. Also, S and F agree on the points (-i, -1), (-k, -1) and (-k, 0) for all $0 \le i \le k - 1$. Consider the seed configuration as shown in Fig.~\ref{imageCompareExample}. According to Lemma 1 of \cite{XuncaiSubDiv}, we will get unique configuration for the comparison tile system which will terminate with CR attaching at (1, 0) position. Now, if $bd_N(CR)$ is $= or >$, then only SSR will attach to this configuration since it is the only matching tile. Now, we have new seed configuration which resemble to subtract and shift seed configuration. Again, according to Lemma 3 of \cite{XuncaiSubDiv}, this will produce unique final configuration in $k + 1$ steps which terminates when SSL tile attaches at (-k, 0). If on the other hand, $bd_N(CR)$ is $<$ than SR tile will attach. In the case also, according to Lemma 2 of \cite{XuncaiSubDiv}, we will get a unique tile configuration in $k + 1$ steps. This process terminate when SL will attach at (-k, 0) position. After either of this configuration is formed, IL will attach at (-k, -1) position since we have $\tau = 2$ which makes it the only possible tile to be attached to this growing assembly. According to Theorem 1, this insertion seed configuration will produce unique output tile assembly. 

Now, the same process repeats unless we get the desired number of bits after decimal place. Therefore, according to induction, the entire configuration will be unique. After all the computation is finished, the corner tiles and boundary tiles will attach at the top to stop the growth of assembly. Hence, square-root computation is terminated $\blacksquare$

\begin{mydef}
The time taken for computing square-root is \BigO{(k_1 + k_2)^2} where $k_1$ is number of bits before decimal and $k_2$ is number of bits after decimal.
\end{mydef}

\textbf{Proof.} In our previous proofs, we have already shown that all the tile assemblies are step configuration which means that only one tile is attached to the seed configuration at a time. Also, since $\tau = 2$ no tile will get attached to the configuration unless it has two surrounding tiles. Also, our tile assembly takes $k_c$ levels to compute and in each level we have $n = k_1 + k _2$ computations. This makes total computation time to be equal to $k_c * (k_1 + k_2)$ where $k_c = 3n/2$. Therefore, since $n = k_1 + k_2$, we have running time complexity as \BigO{(k_1 + k_2)^2}$\blacksquare$

\begin{center}
    \begin{tabular}{| c | c |}
    \hline
    Basic Tiles & Tile Types \\ \hline 
    Left Frame & $2n$ \\ \hline
    Right Frame &  $11$ \\ \hline
    Corner & $4$ \\ \hline
    Top Frame & $4$ \\ \hline
    Input & $4$ \\ \hline
     Computational & $12 + 8 + 16 + 4n$ \\ \hline 
    \end{tabular}
    \captionof{table}{Tile Set to compute square-root of n bit number. The number n is of even length and if not then made even.} \label{tileTable} 
\end{center}

\section{APPLICATIONS}
The method of square-root can be used to compute rational numbers. Also, a slight modification can help us to compute an irrational number $\pi$.

\subsection{COMPUTING RATIONAL NUMBERS}
\begin{figure}[ht!]
\centering
\includegraphics[width=0.5\textwidth]{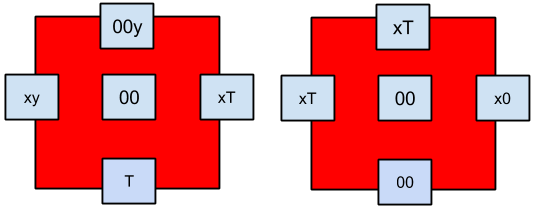}
\caption{Additional 9 computational tiles required for comparing bits. Here x $\epsilon\{<, >, =\}$ and y $\epsilon \{0, 1\}$.}
\label{imageRationalCompare}
\end{figure}

\begin{figure}[ht!]
\centering
\includegraphics[width=0.5\textwidth]{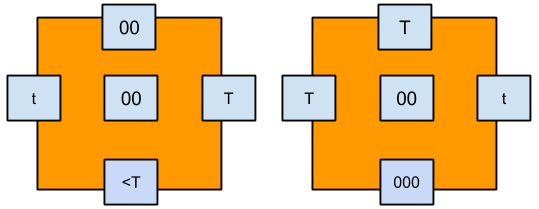}
\caption{Additional 2 computational tiles required for shifting divisor bits by one position.}
\label{imageRationalShift}
\end{figure}

\begin{figure}[ht!]
\centering
\includegraphics[width=0.5\textwidth]{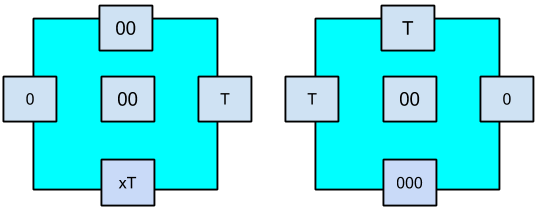}
\caption{Additional 2 computational tiles required for subtracting and shifting bits. Here, x $\epsilon\{ >, =\}$.}
\label{imageRationalSubShift}
\end{figure}

\begin{figure*}[ht]
\includegraphics[scale=0.3]{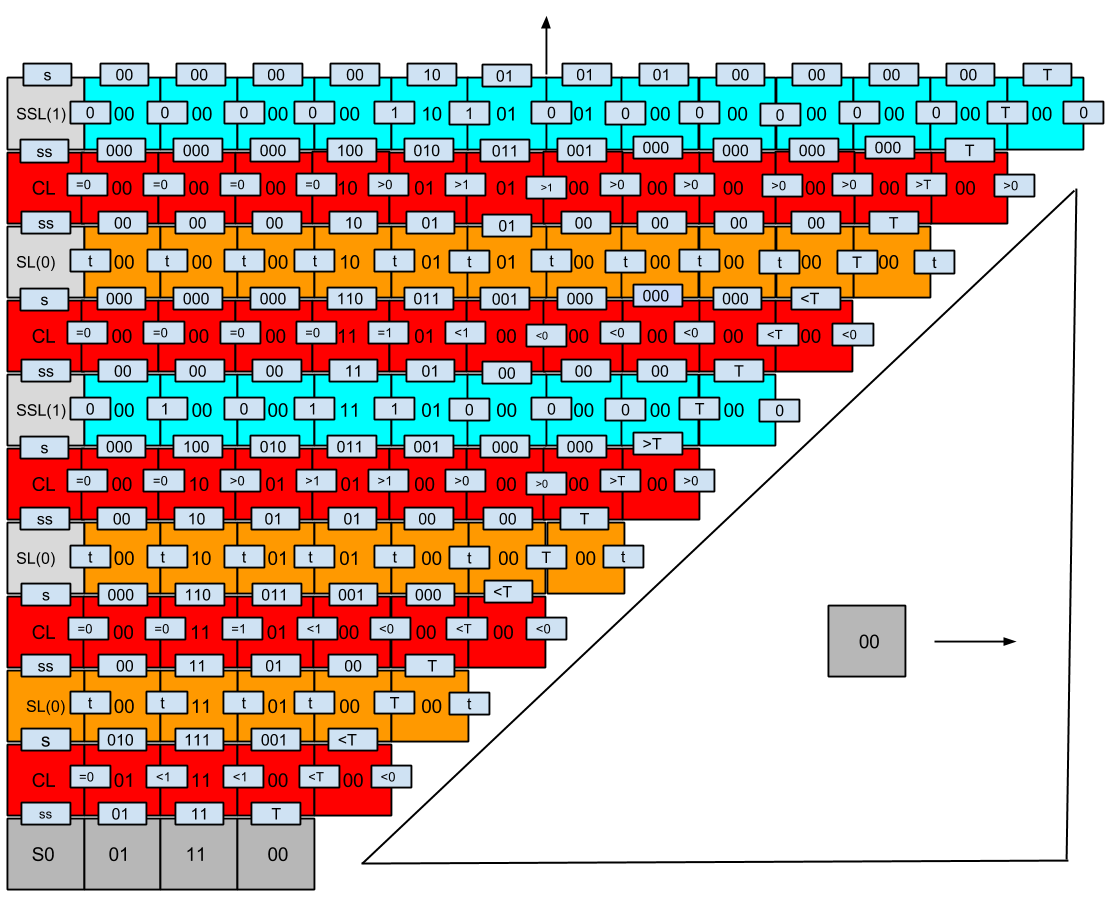}
\caption{Here we divide p = 1 by q = 3 to calculate the rational number 0.3333$\ldots$. Here p is encoded in binary as $10_{2}$ and q is encoded as $11_2$. We get $0.0101.._{2}$ as the output at the left side of the tile assembly model. Arrows indicate the tile assembly is unbounded while triangle indicates that the entire space will be filled with (0, 0) tiles with relevant glues.}
\label{imageInfiniteGrowth}
\end{figure*}

Suppose we are interested in finding the square-root of 1/3 then, first, we need it in decimal form. Therefore, we propose a method to compute rational number p/q given two input integers p and q. It is an infinitely growing tile assembly which is bounded from west and south, the other ends being open. The method to compute rational number requires some new additional computational tiles but, fortunately, this method will not require the boundary tiles. Fig.~\ref{imageRationalCompare},~\ref{imageRationalShift}, and~\ref{imageRationalSubShift} shows additional tiles required to make an infinite assembly growth. The growth for computing 1/3 is shown in Fig.~\ref{imageInfiniteGrowth}.

\begin{mydef}
Let $\sum$ be an alphabet defined as

\begin{tabular}{ l l r }
  $\sum =$  & $\{0, 1, 00, 01, 10, 11, T, t,  <, >, = \}$ & $\bigcup$ \\
  & \{xy $|$ x $\epsilon\{<, >, =\};$ y $\epsilon \{T, 00, 01, 10, 11\}\}$ & $\bigcup$ \\
  & $\{000, 001, 010,  011, 100, 101, 110, 111\}$ &  \\
\end{tabular}
\\\.\\and $T_3$ be the tile set defined over $\sum$. See Table \ref{tileTable}, given that $\tau = 2$ and g=1, can compute a rational number from a given numerator and denominator.Therefore, the tile assembly is defined as defined as  $S_R = (T_3, S, g, \tau)$ where S is the seed configuration.. Then, there exists some $(x_0, y_0) \epsilon Z_2$ such that : $S_R(x_0 - k, y_0 -1) = S_0$, $S_R(x_0, y_0 - 1) = 00$, $S_R(x_0 - k, y_0) = CL$, and for all i $\epsilon \{0, 1,...,k-1\}$, $bd_N(x_0 - i, y_0 - 1) = pq$. Then, the configuration S can produce a unique and unbounded final configuration F and will compute rational number.
\end{mydef}

\textbf{Proof.} Consider the tile system $S_R$. Let input tiles be T =
\\ \{$D_{00} = \{00, null, null, null\}, S_L = \{null, =ss, null\} \\ 
D_{01} = \{01, null, null, null\}, D_{10} = \{10, null, null, null\}, \\
D_{11} = \{11, null, null, null\}, S_0 = \{\$ix, null, null, null\}, \\
00 = \{T, null, null, null\}$\}. \\

Let the seed configuration $S: Z_2 \rightarrow T$ be:
\begin{equation}
 \begin{cases}
    00 = S_R(1, -1)\\
    S_0 = S_R(-k, -1)\\
    CL = S_R(-k, 0)\\
    \forall i \epsilon \{0, 1, .., k-1\}, S_R(-i, -1) = D_{pq}\\
    $For all other (x, y)$ \epsilon Z_2, S_R(x, y) = empty
  \end{cases}
\end{equation}

Let the final configuration be F. Also, S and F agree on the points (-i, -1), (-k, -1) and (-k, 0) for all $0 \le i \le k - 1$. The seed configuration $S$ for the tile growing tile assembly shown in Fig.~\ref{imageInsertExample} resembles the original compare, shift, and  subtract/ shift configurations. There is only on position where the incoming tile can attach but, in this case, after the terminating tile has attached we can have growth on both of its sides. Left side of the terminating tile is exactly similar to division seed configuration hence, according to Lemma 4 of \cite{XuncaiSubDiv}, it will have unique growth. As far as the growth of right side of terminating tile is concerned, there are only zero tiles with all the glue colors having values like $\{000, 00, 0, <0, >0, =0\}$ depending on their level. 

The same process continues perpetually. Therefore, according to induction, the entire configuration will be unique. Therefore, we get an infinite recurring number. If the rational number is terminating then we will get trailing zeroes. 

\begin{center}
	\label{tileTable} 
    \begin{tabular}{| l | l |}
    \hline
    Type of tiles & Number of Tiles \\ \hline 
    Compare & 12 + 9 = 21 \\ \hline 
    Shift & 8 + 2 = 10 \\ \hline 
    Subtract \& Shift & 16 + 2 = 18 \\ \hline 
    Left Boundary & 3\\ \hline
    \end{tabular}
    \captionof{table}{Tile Set to compute rational number $p/q$.} 
\end{center}

\subsection{COMPUTING IRRATIONAL NUMBER - PI($\pi$)}
In the final section of the paper, we will demonstrate a way to compute $\pi$ using an infinite series. The computation method uses same tile systems - Subtraction, Shift, Duplicator, Insertion - to compute the value of an infinite series. In order to compute the value of $\pi$, we have used $Gregory–Leibniz$ series which is defined as: 
\begin{center}
{\Large$\sum\limits_{n=0}^\infty {\frac{(-1)^n}{2n + 1}}= \frac{\pi}{4} = (1 - \frac{1}{3} + \frac{1}{5} - \frac{1}{7} + \ldots)$}
\end{center}
To compute this series, we will require values of all the fractions so that we sum them up. To find the value of fractions, therefore, we can use Theorem 4 which can compute the value of fraction. Once we have the number in decimal form, we simply need to do addition or subtraction to already existing series sum.

\begin{figure*}
\includegraphics[scale=0.3]{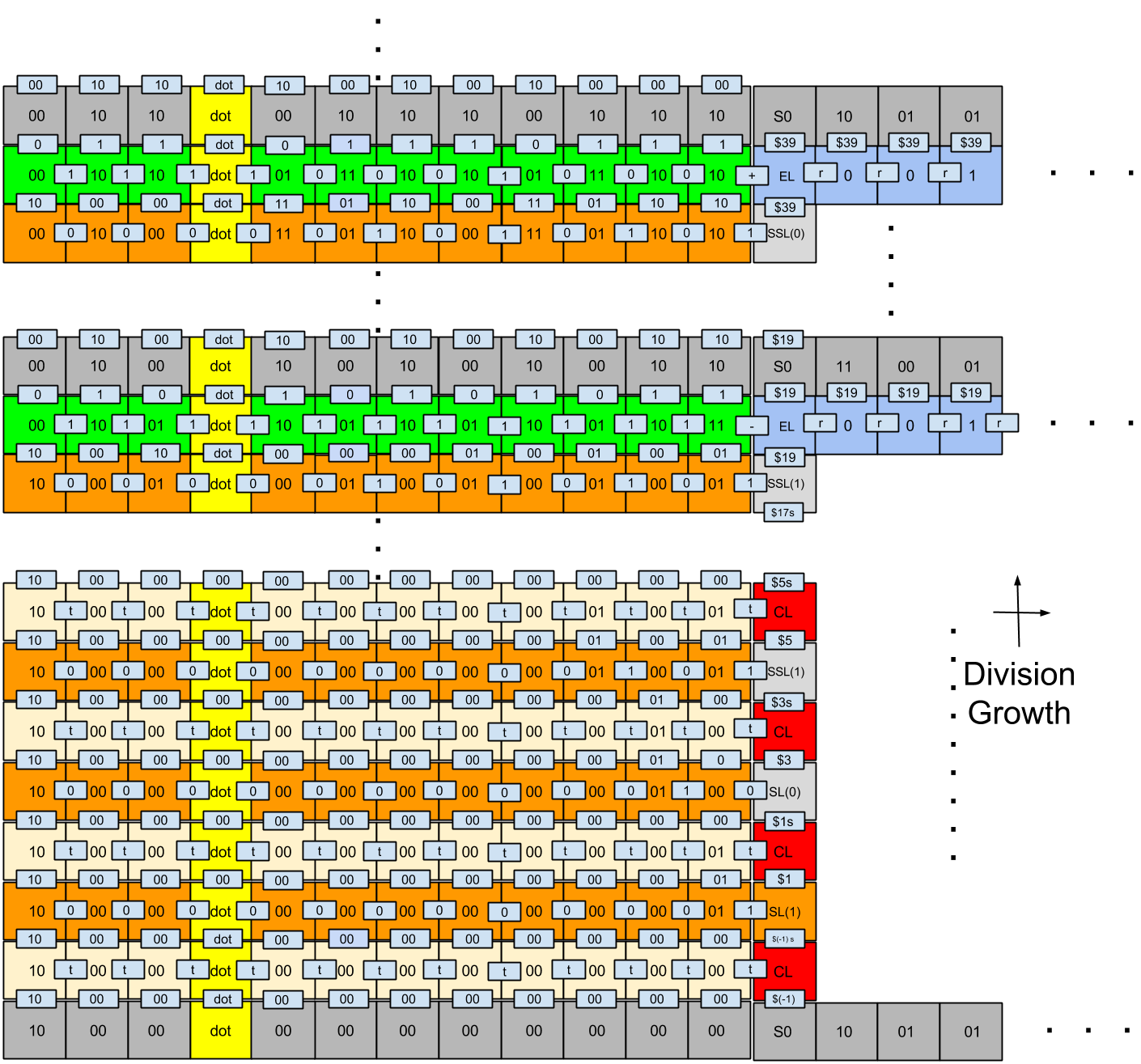}
\caption{Computing value of an irrational number $\pi$ - 3.141592$\ldots$. Here, grey tiles are input tiles, green are addition/subtraction tiles, yellow are copy/duplicate tiles, and orange tiles are shift tiles. Blue tiles indicate the reminder of the corresponding division process.}
\label{imagePI}
\end{figure*}

We need the value of output of division in simpler form so that we can use it for further computation. To do this, we have used copy and shift tile systems. Refer Fig.~\ref{imagePI}. The value of quotient on the left boundary of division tile set is extracted by shifting them one bit per level - yellow and orange tiles - as shown in Fig.~\ref{imagePI}. As a result, when we finish the division process, we have the output value along with existing sum to which we either want to add or subtract this fraction. Now, depending on whether the term is odd or even we need to add or subtract it to the existing sum which is shown by green tiles in Fig.~\ref{imagePI}. This arithmetic occurs meanwhile we have input tiles - grey colored - for next division being attached on the other side. Once this addition or subtraction is finished, we replace the current sum with this new sum in the grey tile as shown in Fig.~\ref{imagePI}. This process keeps run in perpetuity. Depending on the number of precision bits we can control the growth of our assembly.

\begin{mydef}
Let $\sum$ be an alphabet defined as\\
\begin{tabular}{ l l r }
  $\sum =$  & $\{0, 1, 00, 01, 10, 11, T, t,  <, >, =, +, - \}$ & $\bigcup$ \\
  & \{xy $|$ x $\epsilon\{<, >, =\};$ y $\epsilon \{T, 00, 01, 10, 11\}\}$ & $\bigcup$ \\
  & $\{000, 001, 010,  011, 100, 101, 110, 111\}$ & $\bigcup$ \\
  & $\{\$i  | -1 \le i \le n - 1$; $i = 2k + 1\}$ & $\bigcup$ \\
  & $\{\#i x | -1 \le i \le n - 1 ;  x \in \{0, 1, s, ss\}\}$  \\
\end{tabular}
\\\.\\and $T_4$ be the tile set defined over $\sum$. The tile set includes compare, insert, shift, and subtract \& shift tile systems. Given that $\tau = 2$ and g=1, can compute an irrational number $\pi$.Therefore, the tile assembly is defined as defined as  $S_R = (T_4, S, g, \tau)$ where S is the seed configuration.. Then, there exists some $(x_0, y_0) \epsilon Z_2$ such that : $S_R(x_0 , y_0) = S_0$,  $S_R(x_0, y_0+1) = CL$, and for all i $\epsilon \{0, 1,...,k-1\}$, $bd_N(x_0 - i, y_0 - 1) = pq$. Then, the configuration S can produce a unique and unbounded final configuration F and will compute irrational number - $\pi$.
\end{mydef}

\textbf{Proof.} Since we have used the same tile systems - compare, shift, subtract and shift, and duplicator - its proof will be similar to previous theorems. Therefore, we will not go into details of the proof $\blacksquare$\\

There are a few observations about $Gregory–Leibniz$ series~\cite{PIBook}. 
\begin{itemize}
  \item The series converges at a very slow rate~\cite{PIBook}; Leibniz's formula converges extremely slowly: it exhibits sublinear convergence. Calculating $\pi$ to 10 correct decimal places using direct summation of the series requires about five billion terms. 
  \item The series is easy to compute. The numerator of each term is 4 while the denominator increases by 2 at every-step. 
  \item The series does not require multiplication which means we do not require any new tile system.
\end{itemize}
Note that the method to compute $\pi$ may not be most optimized version. As a result, it is an open problem to compute value of $\pi$ in an optimized fashion.

\section{SUPPLEMENTARY FILE}
We have also written a program to generate .tile file of square-root which can be downloaded from our website http://www.guptalab.org.

\section{CONCLUSION}
Our method to compute square-root, and rational numbers is simply a modification of the Xuncai et. al. method to find subtraction and division, using DNA tiles. Our system uses \BigO{n} computational tiles for square-root of an n bit binary number and \BigO{1} tiles for computing rational number. However, it would be an interesting to compute $\pi$ using finite number of tiles.

\bibliography{references}
%
%

\end{document}